\documentclass[aps,preprint,a4paper,showpacs,showkeys,superscriptaddress,nofootinbib]{revtex4-1}
\usepackage{latexsym}
\usepackage{amsmath}
\usepackage{amssymb}
\usepackage{graphicx}
\usepackage{subfigure}
\usepackage{xcolor}
\usepackage{cancel}
\usepackage{multirow,tabularx}%{boldline,array}
\usepackage{hyperref}     % color hypertext for pdflatex
\usepackage[capitalise]{cleveref}

\newcolumntype{C}[1]{>{\centering\arraybackslash}p{#1}}
\hypersetup{colorlinks,%
  citecolor=blue,%
  linkcolor=cyan}%
\usepackage[titletoc]{appendix}
\DeclareMathOperator{\sech}{sech}

\begin{document}
\count\footins = 1000
%\preprint{arXiv:yymm.nnnn [gr-qc]}

\title{On thermal radiation of de Sitter space
in the semiclassical Jackiw-Teitelboim model}

\author{Hwajin Eom}%
\email[]{um16@sogang.ac.kr}%
\affiliation{Department of Physics, Sogang University, Seoul, 04107,
  Republic of Korea}%
\affiliation{Center for Quantum Spacetime, Sogang University, Seoul 04107, Republic of Korea}%

\author{Wontae Kim\footnote[1]{Corresponding author}}%
\email[]{wtkim@sogang.ac.kr}%
\affiliation{Department of Physics, Sogang University, Seoul, 04107,
  Republic of Korea}%
\affiliation{Center for Quantum Spacetime, Sogang University, Seoul 04107, Republic of Korea}%

\date{\today}

\begin{abstract}
In general, the Gibbons-Hawking temperature based on the Euclidean functional approach shows that
de Sitter space in the Bunch-Davies vacuum is globally thermal.
In the exactly soluble semiclassical Jackiw-Teitelboim model, we investigate thermal property of de Sitter space
by taking into account the quantum back reaction of the geometry.
The proper temperature of de Sitter space in the Bunch-Davies vacuum is found to vanish.
In case of a certain quantum state breaking the de Sitter symmetry, de Sitter space can be made thermally exited;
however, in this case the dilaton singularity cannot be avoided.
Consequently, in the Jackiw-Teitelboim model
the proper temperature of de Sitter space in the Bunch-Davies vacuum turns out to be zero and
the Bunch-Davies vacuum is found to be the only physical vacuum without any naked singularities.
\end{abstract}

%

% \pacs{04.70.Dy, 04.62.+v, 04.60.Kz }

\keywords{de Sitter space, Bunch-Davies vacuum, Gibbons-Hawking temperature, proper temperature, Jackiw-Teitelboim model}

\maketitle
%%%%%%%%%%%%%%
%% RevTeX Style End %%
%%%%%%%%%%%%%%

%\newcommand{\lp}{\ell_P}

\newpage
\section{introduction}
\label{sec:introduction}
Since a spacetime during inflation can be approximated as de Sitter space
of a vacuum solution of the Einstein equation with a positive cosmological constant,
de Sitter space has been extensively studied in cosmology.
%an apoch of accelerated expansion of our universe.
For example, the slow-roll inflation \cite{Linde:1981mu,Albrecht:1982wi,Linde:1983gd,Ade:2015lrj}
is typically assumed to begin with the quantum state describing perturbations
in the Bunch-Davies vacuum as
a de Sitter-invariant vacuum of de Sitter space \cite{Nelson:2012ax}.
In fact, Hawking radiation in de Sitter phase of an inflationary
universe is the source of primordial matter fluctuations and classical energy-density perturbations that may
explain the origin of galaxies (for a review, see Ref.~\cite{Brandenberger:1984cz}).

Using the Euclidean functional approach, Gibbons and Hawking \cite{Gibbons:1977mu}
showed that for freely falling observers
the propagator for a scalar field was turned out to be a periodic function of a time with period $i \beta$.
This is the characteristic of thermal Green's functions with a temperature $T=\beta^{-1}$.
In thermal equilibrium the temperature is
identified with the surface gravity at the cosmological horizon as
\begin{equation}
\label{ht}
T_{\rm GH} = \frac{H}{2\pi},
\end{equation}
where $H$ is the Hubble constant.
It should be emphasized that the Gibbons and Hawking's formulation is based on
the belief
that a well-defined meaning for the Green's function can be given by analytically continuing
back to the original Lorentzian spacetime
from the Euclidean metric.

On the other hand, in the stress tensor approach, it has been known that
the temperature in thermal equilibrium can be derived from the Stefan-Boltzmann law \cite{Tolman:1930zza,Tolman:1930ona}.
However, in two-dimensional de Sitter space,
this procedure fails since
one immediately confronts a negative energy density as
\begin{equation}
  \epsilon= -\langle T^0_0 \rangle_{\rm BD}=-\frac{H^2}{24 \pi} =\gamma_{(2)} T^2_{\rm GH}, \label{2d}
\end{equation}
where $\gamma_{(2)}=\pi/6$ for a single massless scalar field.
This fact is not new since
the negative sign of the energy density
was already pointed out in Ref.~\cite{Markkanen:2017abw}.
The usual Stefan-Boltzmann law
renders the Gibbons-Hawking temperature to be imaginary, so
it does not meet the stress tensor approach.
Now, in the stress tensor approach, it raises a question about
how to define the proper temperature of two-dimensional de Sitter space
in thermal equilibrium.
Thus, we are going to study this issue both on the de Sitter background
and from the dynamical point of view,
in the semiclassically quantized Jackiw-Teitelboim model with a positive cosmological constant \cite{Teitelboim:1983ux,Jackiw:1984je}, respectively.

In thermal equilibrium, it is worth noting that Tolman derived the Stefan-Boltzmann on the curved spacetime by assuming that
the stress tensor is {\it traceless} \cite{Tolman:1930zza,Tolman:1930ona}; however, this fact was sometimes ignored.
In this regard,
the Stefan-Boltzmann law for the non-vanishing trace of the stress tensor should be
modified in such a way that the trace anomaly of the stress tensor must be appropriately taken into account
in the thermodynamic first law \cite{Gim:2015era}.
In fact, for a non-vanishing trace of the stress tensor, the modified Stefan-Boltzmann law was successfully applied to
a cosmological model for reheating of our universe \cite{Gim:2016uvv} and black hole models
related to quantum atmosphere \cite{Unruh:1977ga,Giddings:2015uzr,Eune:2015xvx,Kim:2016iyf,Eune:2017iab,Eune:2019aat}.
Since the trace of stress tensor on the de Sitter space is also non-vanishing \cite{Davies:1977ze,10.2307/79520,Christensen:1977jc,Deser:1976yx},
we shall apply this modified Stefan-Boltzmann law to de Sitter space in the stress tensor approach,
and will resolve the imaginary temperature problem in Eq.~\eqref{2d}.

The organization of this paper is as follows.
In Sec.~\ref{sec:review_de Sitter2}, we will calculate the proper temperature of
de Sitter space by using the renormalized stress tensor from the one-loop Polyakov effective action of free scalar fields
and obtain the vanishing proper temperature in the Bunch-Davies vacuum by using the modified Stefan-Boltzmann law.
In Sec.~\ref{sec:JT}, we will revisit thermal property of de Sitter space
by embedding de Sitter space in the semiclassically quantized Jackiw-Teitelboim model, and
obtain the same result as that of Sec.~\ref{sec:JT}:
the proper temperature in the Bunch-Davies vacuum vanishes.
Of course, in case of de Sitter non-invariant quantum states de Sitter space can be made thermally exited;
however, a naked singularity appears.
Finally, conclusion and discussion will
be given in Sec.~\ref{sec:conclusion}.

\section{On the Background of De Sitter space}
\label{sec:review_de Sitter2}
We start with the length element of de Sitter space described by
(for a review, see Ref.~\cite{Spradlin:2001pw})
\begin{equation}
    \textrm d s^2 = -g(r)\textrm d t^2 +\frac{1}{g(r)} \textrm d r^2,
\label{eq:metric}
\end{equation}
where $g(r) = 1-H^2 r^2$.
Note that the cosmological horizon is defined as
$r_H=1/H$ and the spacetime is locally flat at $r=0$.
By using the tortoise coordinate
$r^* = r_H \tanh^{-1} (H r)$,
we rewrite the length element \eqref{eq:metric} in terms of conformal coordinates as
\begin{eqnarray}
\label{eq:metric_lightcone}
    \textrm d s^2 &=& - e^{2 \rho(\sigma^+,\sigma^-)} \textrm d \sigma^+ \textrm d \sigma^-,  \\
     e^{2 \rho (\sigma^+,\sigma^-)} &=& \sech^2 \left[ \frac{H}{2}  (\sigma^+ -\sigma^-)\right], \nonumber
    \end{eqnarray}
where $\sigma^\pm=t\pm r^*$.
Although two-dimensional de Sitter space covers
$-\infty < t <\infty$  and $0\leq |r| <H^{-1}$,
we will restrict the range of the radial coordinate to $0\leq r <H^{-1}$.
Thus, the Kruskal coordinates are obtained as $x^\pm =\mp e^{\mp H \sigma^\pm}/H$.
%\begin{equation}
 %   x^\pm =\mp \frac{1}{H} e^{\mp H \sigma^\pm} =\mp \frac{e^{\mp Ht}}{H}\sqrt{\frac{1-Hr}{1+Hr}},
 %   \label{eq:Kruskal}
%\end{equation}
%we can also obtain the Kruskal form of the length element as
%\begin{equation}
%\label{eq:metric_global_lightcone}
%    \textrm d s^2 = -e^{2 \tilde{\rho}(x^+,x^-)} \textrm d x^+ \textrm d x^-,
%        e^{2 \tilde{\rho}(x^+,x^-)}=\frac{4}{(H^2 x^+ x^- -1)^2}.
%\end{equation}
%where our spacetime of our interest is shown in Fig.\ref{fig:Penrose}.

Next, we consider
the renormalized stress tensor from the one-loop Polyakov effective action of free scalar fields \cite{Polyakov:1987zb},
which can also be determined by solving the covariant conservation law of stress tensor and
the trace-anomaly relation as \cite{Christensen:1977jc}
\begin{eqnarray}
\label{eq:em tensor}
	\langle T_{\pm\pm} (\sigma) \rangle &=& -\frac{\kappa}{8\pi}  \left[ \frac 1 2 (g')^2-g g'' \right]-\frac{\kappa}{\pi} t_\pm (\sigma^\pm) ,\\
     \langle T_{+-} (\sigma) \rangle &=& \frac{\kappa}{8\pi} g g'',\label{tensor2}
\end{eqnarray}
where the above prime denotes the derivative with respect to $r$. $t_\pm (\sigma^\pm)$ are integration functions,
and $\kappa=N/12$ where $N$ is the number of classical matter fields.
Plugging the metric function \eqref{eq:metric} into Eqs.~\eqref{eq:em tensor} and \eqref{tensor2},
we can write the explicit form of the stress tensor.
%\begin{equation}
%\label{eq:em tensor ds2}
%	\langle T_{\pm\pm} (\sigma) \rangle = -\kappa \left[\frac{H^2}{4}+t_\pm (\sigma^\pm) \right],
%    \qquad \langle T_{+-} (\sigma) \rangle = -\frac{\kappa H^2}{4} (1-H^2 r^2).
%\end{equation}

From the definition of the Bunch-Davies vacuum $|0 \rangle_{\rm BD}$
which states that
the stress tensor be regular at both the future and past horizons \cite{Dowker:1975tf,birrell1984quantum},
we can determine
\begin{equation}
\label{eq:BD vacuum}
t_\pm (\sigma^\pm) =-\frac{H^2}{4},
\end{equation}
and thus, the vacuum expectation value of the stress tensor
\eqref{eq:em tensor} and \eqref{tensor2} becomes
\begin{equation}
\label{eq:em tensor in BD}
    \langle T_{\mu\nu} \rangle_{\rm BD} = \frac{\kappa H^2}{2 \pi} g_{\mu\nu}
\end{equation}
which is de Sitter-invariant.

Now, one might ask in two-dimensional de Sitter space how to get the proper temperature from the negative energy density.
Let us first remind of the conventional process.
We are aware of the fact that
both the influx and outward flux must vanish;
in other words, $\langle T_{\pm\pm}  \rangle_{\rm BD}=0$
as seen from Eq.~\eqref{eq:em tensor in BD}.
Thus, the only source contributing to the energy density must be the off-diagonal component of the stress tensor.
%\section{without quantum back reaction}
%\label{sec:temp}
%In this section,
%we introduce the conventional temperature derived
%from the Stefan-Boltzmann law in the Bunch-Davies vacuum,
%and then, point out an issue upon the local temperature suffering from the imaginary value problem.
%Finally, we resolve this issue by using a modified Stefan-Boltzmann law.
%In order to obtain the local temperature of the two-dimensional de Sitter space in
%thermal equilibrium, let us start with the Stefan-Boltzmann law for gravitational systems defined as %\cite{Tolman:1930zza,Tolman:1930ona}
%\begin{equation}
%\label{eq:sb law}
%\rho = \gamma  T^2,
%\end{equation}
%where $\rho$ is the proper energy density with the Stefan-Boltzmann constant for scalar fields $\gamma =N(\pi/6)$.
Explicitly,
the proper energy density for a local observer is
\begin{equation}\label{com}
\epsilon=\langle T_{\mu\nu}\rangle u^\mu u^\nu= \frac 1 g \bigg[\langle T_{++} \rangle + \langle T_{--} \rangle+2 \langle T_{+-} \rangle \bigg]
\end{equation}
with the two-velocity $u^\mu =(u^+ (\sigma), u^- (\sigma)) = (1/\sqrt g)(1,1)$
satisfying $u^\mu u_\mu=-1$.
The energy density in the Bunch-Davies vacuum
is related to the proper temperature via the Stefan-Boltzmann law of
$\gamma T^2 =\epsilon$ \cite{Tolman:1930zza,Tolman:1930ona} as
%\begin{equation}
%\label{rho}
%    \rho =  \frac 1 g \bigg[\langle T_{++}\rangle +\langle T_{--}\rangle+2 \langle T_{+-}\rangle \bigg].
%\end{equation}
$\gamma  T^2 =\left(1/g\right)\left[\langle T_{++} \rangle_{\rm BD}+\langle
T_{--} \rangle_{\rm BD}+2 \langle T_{+-} \rangle_{\rm BD} \right]$.
Hence, the temperature in the Bunch-Davies vacuum can be calculated as
\begin{equation}
  \gamma T^2= -\frac{\kappa H^2}{2\pi} \label{SB}
\end{equation}
which is the same expression as Eq.~\eqref{2d} apart from $N$ where $\gamma=N(\pi/6)$ for $N$-scalars.
If we take the absolute value of the proper energy density such as
$ \gamma T^2=|\epsilon | $,
the two-dimensional Gibbons-Hawking temperature may be obtained as $T_{\rm GH} = H/(2\pi)$  \cite{Gibbons:1977mu};
however, such an {\it ad-hoc} process seems to be unwarranted.

%as is seen from Eq.~\eqref{SB}.

It should be emphasized that Tolman derived the Stefan-Boltzmann in the curved spacetime by assuming that
the stress tensor is traceless \cite{Tolman:1930zza,Tolman:1930ona}.
Thus, the Stefan-Boltzmann law should be
modified when the stress tensor is not trace free because of conformal anomaly \cite{Gim:2015era}.
The modified Stefan-Boltzmann law takes the form of \cite{Gim:2015era}
\begin{equation}
\label{eq:eff sb law}
    \epsilon = \gamma  T^2 -\frac 1 2 \langle T^\mu_\mu \rangle.
\end{equation}
In the limit of the traceless stress tensor, this modified Stefan-Boltzmann law
reduces to the usual Stefan-Boltzmann law.
From Eq.~\eqref{eq:eff sb law}, the proper temperature of
thermal radiation in the Bunch-Davies vacuum can be written as
\begin{equation}
\label{newcom}
\gamma  T^2 = \frac 1 g \bigg[\langle T_{++}  \rangle_{\rm BD} +\langle  T_{--} \rangle_{\rm BD} \bigg]
\end{equation}
which tells us that the origin of the proper temperature is just the ingoing and outgoing fluxes
dropping the off-diagonal component of the stress tensor in Eq.~\eqref{com}.
From Eq.~\eqref{eq:em tensor in BD} in the Bunch-Davies vacuum, the proper temperature \eqref{newcom} naturally becomes
\begin{equation}
\label{eq:nonthermal}
T=0,
\end{equation}
which is a
plausible result in the sense that the cosmological constant plays a role of the potential energy, so
it is of no relevance to kinetic excitations of particles \cite{Gim:2016uvv}.
Therefore, the proper temperature of de Sitter space vanishes in the Bunch-Davies vacuum.%see the same vanishing proper temperature.

As a comment, is it impossible to  get thermal de Sitter space?
If the boundary condition $t_{\pm}(\sigma^\pm)$ is chosen for a quantum state $| \Psi \rangle$ as
\begin{equation}
\label{eq:Hawking temp vacuum}
t_\pm (\sigma^\pm) =-\frac{H^2}{2},
\end{equation}
then the stress tensors are obtained as
\begin{equation}
\langle \Psi| T_{\pm\pm} (\sigma)|\Psi \rangle =\frac{\kappa H^2}{4 \pi},\quad
\langle \Psi | T_{+-}(\sigma)| \Psi \rangle = -\frac{\kappa H^2}{4 \pi} g(r),
%\langle T_{\mu\nu} (\sigma) \rangle_{\rm H} = \frac{\kappa H^2}{2} g_{\mu\nu} -\frac{\kappa H^2}{4} I_{\mu\nu}
\end{equation}
where the de Sitter symmetry of the stress tensor is broken.
From Eq.~\eqref{newcom}, the proper temperature can be read off as
\begin{equation}
\label{Tolman}
T = \frac{T_H}{\sqrt{g(r)}}.
\end{equation}
It means that if we choose de Sitter non-invariant vacuum, then de Sitter space can be
thermal but the proper temperature in this quantum state is observer-dependent and singular at the cosmological horizon.
This case will be ruled out in the next section by the singular behavior of the dilaton field.
As a result, as long as we choose the Bunch-Davies vacuum,
the proper temperature should vanish globally like Eq.~\eqref{eq:nonthermal}.

\section{De Sitter SPACE IN the JACKIW-TEITELBOIM MODEL}
\label{sec:JT}
In this section, we study thermal behavior of de Sitter space by using the semi-classical Jackiw-Teitelboim (JT) gravity
in order to find out whether the result \eqref{eq:nonthermal} persists or not
when taking into account the quantum back reaction of the geometry.
Because the JT model is
one of the simplest models in two-dimensional dilaton gravity,
it has been studied in variety of cases of interest \cite{Mann:1989gh,Muta:1992xw,Lemos:1996bq,Grumiller:2014oha,Cvetic:2016eiv,Kitaev:2017awl,Almheiri:2019qdq,Momeni:2020tyt}.
It has also been investigated recently in the context of the generalized model \cite{Almheiri:2014cka,Engelsoy:2016xyb,Anninos:2017hhn,Grumiller:2021cwg}
where the generalized dilaton potential $\mathcal V (\phi,-(\partial \phi)^2)$ is employed.
In this generalized model,
the JT model corresponds to the case of $\mathcal V = \Lambda \phi$ with a constant $\Lambda$.
This form of the dilaton potential $\mathcal V$ is naturally motivated by
the dimensional reduction of the Einstein-Hilbert action
with spherical symmetry,
which shows that the dilaton field couples linearly to both the scalar curvature and
cosmological constant.
That is, the constant $\Lambda$ plays the role of the cosmological constant in
the Einstein-Hilbert action
in the sense that it gives the constant scalar curvature
corresponding to anti-de Sitter and de Sitter space for
the negative and positive $\Lambda$, respectively \cite{Grumiller:2014oha}.
In this regard, the JT model could be regarded as an dimensionally reduced effective theory,
which describes de Sitter space exactly.

%the bulk action in the generalized dilaton gravity in two dimensions
%$S_{\rm 2d} \propto \int d^2 x \sqrt{-g} \left[\phi R-2\mathcal V (\phi,-(\partial\phi)^2)\right]$

Let us start with the JT model described by the action \cite{Teitelboim:1983ux,Jackiw:1984je}
\begin{equation}
\label{eq:action for JT}
S_{\rm JT}=\frac{1}{2 \pi}\int d^2 x \sqrt{-g}\phi \left[R -2H^2 \right],
\end{equation}
where $\phi$ is a dilaton field to implement a constraint for constant scalar curvature $2H^2$.
The classical  and quantum effective actions
are written as
\begin{eqnarray}
    S_{{\rm{cl}}} &=&  \frac{1}{2 \pi} \int\/d^{2}x \sqrt{-g}\,
       \sum_{i=1}^{N}\left[ - \frac12  (\nabla f_i)^2 \right],
\label{cm}\\
    S_{{\rm{qt}}} &=& -\frac{\kappa}{8 \pi}\int d^2 x \sqrt{-g}
    \left[ R\frac{1}{\Box}R \right],
\label{qm}
\end{eqnarray}
respectively, where $R$ is the scalar curvature, $f_i$ are classical scalar fields,
$\Box$ is the covariant d'Alembert operator defined by $\Box=\nabla^\mu \nabla_\mu$
and Eq.~\eqref{qm} is the Polyakov action
\cite{Polyakov:1987zb}.
We will consider the large $N$ limit in the semiclassical regime
in order to make a full sense of semiclassical equation of motion, where the other quantum corrections from
the dilaton gravity sector can be neglected.

Restoring $\hbar$, we note that $\kappa=\hbar N/12$ in Eq.~\eqref{qm}.
In the large $N$ limit with $N\phi^{-1}$ held fixed,
the JT action \eqref{eq:action for JT} and the quantum effective action \eqref{qm} are the same leading order
because the dilaton field $\phi$ playing a role of inverse of coupling  is order $N^1$.
However, the classical limit of the total action can be achieved by taking the limit of $\hbar\to 0$,
$i.e.$, $\kappa\to 0$, which leads to the classical JT action.

By introducing an additional auxiliary field $\psi$, the localized action for Eq.~\eqref{qm}
can be written for convenience as
\begin{equation}
    S_{\rm qt} = -\frac{\kappa}{8 \pi} \int d^2 x \sqrt{-g}
    \left[\left(\nabla\psi\right)^2 -2 R\psi \right].
\end{equation}
%From the total action $S_{\rm tot} = S_{\rm JT}+S_{\rm cl}+S_{\rm qt}$,
%the equation of motion for the metric tensor is given by
%\begin{align}
%    \notag
%    \frac{2\pi}{\sqrt{-g}} \frac{\delta S_{\rm tot}}{\delta g^{\mu\nu}}
%    =~ & G_{\mu\nu} \phi -\nabla_\mu \nabla_\nu\phi +g_{\mu\nu} \Box \phi+H^2 \phi
%    +\sum_{i=1}^N \left[-\frac 1 2 \nabla_\mu f_i \nabla_\nu f_i
%    +\frac 1 4 g_{\mu\nu} \left(\nabla f_i \right)^2\right] \\\label{eq:var metric}
%    & +\frac{\kappa\pi}{2} \left[ G_{\mu\nu} \psi
%    -\nabla_\mu \nabla_\nu\psi +g_{\mu\nu} \Box \psi
%    -\frac 1 2 \nabla_\mu \psi \nabla_\nu \psi +\frac 1 4 g_{\mu\nu} \left(\nabla\psi\right)^2 \right] =0,
%\end{align}
%and equations of motion for $\phi$,
%$f_i$, and $\psi$, are written as

From the total action $S_{\rm tot} = S_{\rm JT}+S_{\rm cl}+S_{\rm qt}$
in the conformal gauge \eqref{eq:metric_lightcone},
equations of motion with respect to $g_{\mu\nu}$, $\phi$, $f_i$, and $\psi$
are obtained
as
\begin{align}
\label{eq:metric offdiagonal}
    & \partial_+ \partial_- \phi -\frac{H^2}{2} e^{2\rho} \phi =\pi  T_{+-}(\sigma),\\
    \label{eq:constraint}
    & -\partial^2_{\pm} \phi + 2 \partial_{\pm} \rho \partial_{\pm} \phi  = \pi T_{\pm\pm} (\sigma),\\
    \label{eq:eom rho}
    &  \partial_+ \partial_- \rho - \frac{H^2}{4} e^{2\rho}=0,\\
    \label{eq:eom h}
    &  \partial_+ \partial_- f_i=0,\\
    \label{eq:eom psi}
    & 2 \partial_+ \partial_- \rho -\partial_+ \partial_- \psi = 0,
\end{align}
where the stress tensor for matter is defined as
$T_{\mu\nu}=(-2\delta S)/(\sqrt{-g}\delta g^{\mu\nu})$ which consists of classical and quantum-mechanical parts:
%\cite{doi:10.1142/p378}
\begin{equation}
T_{\mu\nu}=T^{f}_{\mu\nu} + \langle  T_{\mu\nu}  \rangle,
\end{equation}
where
\begin{align}
    T^{f}_{\pm\pm}  &= \frac{1}{2\pi} \sum_{i=1}^N \left(\partial_\pm f_i\right)^2,\quad\left(T_{\rm cl} \right)_{+-}=0, \\
      \langle T_{\pm\pm} \rangle &=
-\frac{\kappa}{2\pi} \left[-\frac 1 2 \left(\partial_\pm \psi\right)^2-\partial_\pm^2 \psi +2 \partial_\pm \rho \partial_\pm \psi\right], \label{crazy}\\
     \langle T_{+-}  \rangle &=  -\frac{\kappa}{2\pi}\partial_+ \partial_- \psi .
     %\langle T_{\pm\pm} (\sigma) \rangle
     %\langle T_{+-} (\sigma) \rangle
    \label{static energy}
\end{align}
For simplicity, we set $f_i=0$ since we are concerned with quantum radiation.
Solving Eq.~\eqref{eq:eom psi}, we can obtain $\psi = 2\rho +\tilde t_+ (\sigma^+) +\tilde t_- (\sigma^-)$,
where $\tilde t_\pm (\sigma^\pm)$ are integration functions.
By eliminating the auxiliary field $\psi$, the quantum stress tensors are written as
\begin{align}\label{eq:stress tensor pp mm}
    &\langle T_{\pm\pm} \rangle=- \frac \kappa \pi \left[ \left(\partial_{\pm} \rho\right)^2- \partial_{\pm}^2\rho\right]   -\frac \kappa \pi  t_{\pm} (\sigma^\pm),\\\label{eq:stress tensor pm}
    &\langle T_{+-} \rangle = - \frac \kappa \pi  \partial_+ \partial_- \rho
\end{align}
after redefinition of integration of functions as $t_\pm=(-1/4)\left\{(\partial_\pm \tilde t_\pm )^2+2 \partial^2_\pm \tilde t_\pm \right\}$.
Note that Eq.~\eqref{eq:constraint} is the constraint equation including $t_{\pm} (\sigma^\pm)$
which reflect the nonlocality of the Polyakov
action.
%\cite{Callan:1992rs}.

Combining Eqs.~\eqref{eq:metric offdiagonal}, \eqref{eq:eom rho} and \eqref{eq:stress tensor pm} gives
\begin{equation}
\label{eq:eom Psi}
    \partial_+ \partial_- \phi + \frac{H^2}{4} e^{2\rho} \left(\kappa-2 \phi\right) =0.
\end{equation}
Solving Eqs.~\eqref{eq:eom rho} and \eqref{eq:eom Psi}, we obtain
\begin{align}
\label{eq:sol rho}
    & e^{2\rho(\sigma^+,\sigma^-)} =  \sech^2 \left[ \frac H 2 \left(\sigma^+ -\sigma^- \right)\right],\\
\label{eq:sol Psi}
    \phi(\sigma^+,\sigma^-)  = \bigg(\!\tilde \alpha + &  \frac{\kappa H}{4} (1-\alpha)\left(\sigma^+ -\sigma^- \right)\!\!\!\bigg)
     \tanh  \left[ \frac{H (\sigma^+ -\sigma^-)}{2}\right]
      +\frac{\kappa \alpha}{2},
\end{align}
where $\alpha$, and $\tilde \alpha$ are integration constants.
Interestingly, the value of the dilation field at the origin
approaches a finite value as $\phi|_{r^*=0} =(\kappa \alpha)/2$, and thus,
the dilaton field evades the singularity at the origin through quantum corrections.
The dynamical equation of motion \eqref{eq:eom rho} is decoupled from the matter source, which renders
the curvature scalar to be constant.

Plugging the metric \eqref{eq:sol rho} into Eqs.~\eqref{eq:stress tensor pp mm} and \eqref{eq:stress tensor pm},
we obtain the stress tensor for quantum radiation as
\begin{align}
\label{totalflux}
    \langle T_{\pm\pm} (\sigma)\rangle  =
-\frac{\kappa H^2}{4\pi}  -\frac \kappa \pi t_{\pm}(\sigma^\pm),\quad
    \langle T_{+-} (\sigma)\rangle = -\frac{\kappa H^2}{4\pi} e^{2 \rho(\sigma^+, \sigma^- )}.
\end{align}
The flux \eqref{totalflux}
should satisfy the constraint equation
\eqref{eq:constraint}, which yields
\begin{equation}
\label{eq:general bc}
     t_\pm (\sigma^\pm) = -\frac{H^2}{4} \alpha,
\end{equation}
where the quantum state of the stress tensor is related to the dilaton configuration.
In Sec.~\ref{sec:review_de Sitter2}, the vacuum state was imposed by hand because we have treated de Sitter space
as the background geometry, but now the Bunch-Davies vacuum state should be determined by the dilaton parameter $\alpha$
thanks to the dynamical treatment of de Sitter space.
From Eqs.~\eqref{totalflux} and \eqref{eq:general bc}, the stress tensor is neatly written as
\begin{equation}
\label{central}
\langle T_{\mu\nu} (\sigma)  \rangle  = \frac{\kappa H^2}{2\pi} g_{\mu\nu}
-\frac{\kappa H^2}{4\pi}(1-\alpha)I_{\mu\nu},
\end{equation}
where the stress tensor is central extended by $I_{\pm\pm}\!=1$ and $I_{\pm\mp}\!=\!0$.
In particular, $\alpha=1$ and $\alpha=2$ correspond to the condition \eqref{eq:BD vacuum} for
the de Sitter-invariant Bunch-Davies vacuum and
the condition \eqref{eq:Hawking temp vacuum} for the de Sitter non-invariant thermal state $| \Psi \rangle$.
So they are written as
\begin{align}
\label{eq:vac alpha1}
\langle  T_{\mu\nu} (\sigma) \rangle_{\rm BD} = & \frac{\kappa H^2}{2\pi} g_{\mu\nu}, \\
\label{eq:vac alpha2}
\langle \Psi |T_{\mu\nu} (\sigma) | \Psi \rangle  = & \frac{\kappa H^2}{2\pi}  g_{\mu\nu}
+\frac{\kappa H^2}{4\pi} I_{\mu\nu}.
\end{align}

\begin{figure}[t]
\centering
\includegraphics[width=0.6\textwidth]{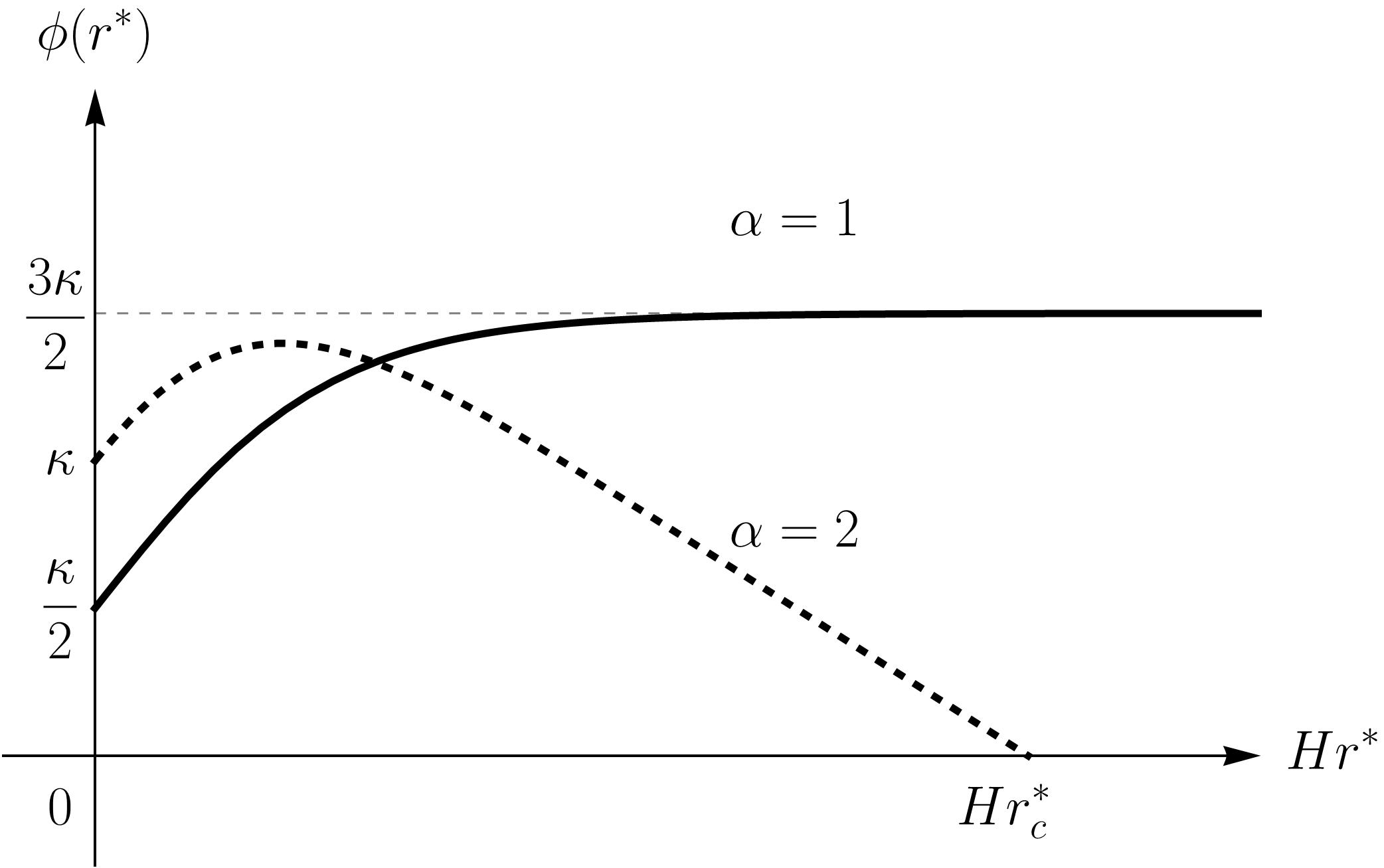}
\caption{The dilaton solution \eqref{eq:sol Psi} is plotted
in the case of $\alpha=1$ and $\alpha=2$,
where $\tilde \alpha=\kappa$ is chosen for simplicity.
Note that the solid  and dashed curves
are dilaton fields for the Bunch-Davies vacuum state ($\alpha =1$)
and for the quantum state ($\alpha=2$), respectively.
The two curves start with positive values of $\kappa/2$ and $\kappa$ at $r^* =0$.
The solid curve converges to a positive and finite value as $r^* \to\infty$ whereas
the dashed curve approaches zero at $r^*=r^*_c$.
%However, a singularity occurs at $r^*=r^*_c$, and so an infinite dilaton coupling happens.
}
\label{fig:dilaton}
\end{figure}

We are now in a position to calculate the proper temperature.
By using the proper energy density \eqref{com},
the modified Stefan-Boltzmann law \eqref{eq:eff sb law}, and
the stress tensor \eqref{central},
the proper temperature for an arbitrary $\alpha$ is calculated as
\begin{equation}
\label{finaltem}
T= \sqrt{\alpha -1}\left(\frac{T_{\rm GH}}{\sqrt g}\right),
\end{equation}
where $\alpha \ge 1$ for a real temperature.
For $\alpha=1$ corresponding to the Bunch-Davies vacuum \eqref{eq:BD vacuum},
the proper temperature vanishes like Eq.~\eqref{eq:nonthermal}.
In case of $\alpha > 1$,
%including $\alpha=2$ which gives $T_H$ at $r=0$,
the redshift factor in the proper temperature \eqref{finaltem} indicates inhomogeneity of the proper temperature
due to the breaking of de Sitter symmetry
except the time-like Killing symmetry implemented by
$\xi^\mu_{(t)} \partial_\mu=\partial_t$.

For the dilaton solution \eqref{eq:sol Psi},
the parameters are shown to satisfy $0 < \alpha \leq 1$
and $\tilde \alpha > -(\kappa \alpha) /2 $ upon requiring the singularity free condition of the dilaton field.
In the special case of $\alpha=2$ of interest,
the dilaton field goes to zero at a finite $r^*$,
so a naked singularity occurs
as seen in Fig.~\ref{fig:dilaton}.
Thus,
both the reality condition for the proper temperature \eqref{finaltem} and
the singularity free condition for the dilaton field
determines $\alpha=1$ uniquely.
As a result,
the de Sitter invariant Bunch-Davies vacuum is the only well-defined vacuum in thermal equilibrium
and the proper temperature globally vanishes in the semiclassical Jackiw-Teitelboim model.

On the other hand,
the JT action \eqref{eq:action for JT} with the replacement
of $H\to i\ell^{-1}$ can describe anti-de Sitter (AdS$_2$) space ($\ell$ : AdS radius),
one might wonder how
solutions in the case of AdS$_2$ space
are related to our solutions \eqref{eq:sol rho} and \eqref{eq:sol Psi}
in de Sitter (dS$_2$) space.
In AdS$_2$ space, it has been known that the dilaton solution $\phi_{\rm AdS}$ diverges at its boundary \cite{Almheiri:2014cka,Engelsoy:2016xyb}.
On the contrary, as seen in Fig.~\ref{fig:dilaton},
the dilaton solution \eqref{eq:sol Psi} in dS$_2$ space
approaches a finite value at the boundary ($r^*\to\infty$).
In fact, a clue to the different behavior of the dilaton solution in AdS$_2$ and dS$_2$ space
at each boundaries
might be found in
the dilaton solution in the generalized models \cite{Anninos:2017hhn,Grumiller:2021cwg}.
In these models,
the solutions ($e.g.$, $e^{2\rho}$ and $\phi$) were
turned out to interpolate between AdS$_2$ and dS$_2$ space.
This fact was clearly shown by the scalar curvature
which approaches either a negative or a positive constant
depending on the large or the small value of dilaton field.

In the CGHS and RST models \cite{Callan:1992rs,Russo:1992ax}, the net flux is always zero in thermal equilibrium
with equal incoming and outgoing fluxes.
The Hartle-Hawking vacuum state is defined as
the state annihilated by the annihilation operators that multiply
the positive frequency modes for both incoming modes in the past horizon and outgoing modes in the
future horizon. It simply means that
there are no incoming and outgoing particles at the horizon.
Despite the absence of excited particles at the horizon, there exist
incoming and outgoing fluxes at infinity because they originates from a macroscopic distance far from the horizon.
This interpretation initiated by Unruh \cite{Unruh:1977ga} is of relevance to the so-called ``quantum atmosphere" recently elaborated by Giddings \cite{Giddings:2015uzr}.
In the limit of asymptotic infinity,
these incoming and outgoing fluxes are responsible for the Hawking temperature.

Let us now point out some differences between the Gibbons-Hawking temperature and the present
proper temperature within the same stress tensor approach in order to compare them evenly.
We define the stress tensor \eqref{eq:stress tensor pp mm} in terms of the bulk stress tensor
and the boundary stress tensor, respectively as
$\langle T_{\pm\pm}\rangle=\langle T_{\pm\pm}^{\rm bulk}\rangle + \langle T_{\pm\pm}^{\rm bdy} \rangle,$
where $\langle T_{\pm\pm}^{\rm bulk} \rangle=
(- \kappa/\pi) \left[ \left(\partial_\pm\rho\right)^2-
 \partial_\pm^2\rho\right]$ and $\langle T_{\pm\pm}^{\rm bdy} \rangle =(- \kappa/\pi) t_\pm (\sigma^\pm)$.
From the boundary stress tensor in the Bunch-Davies vacuum state, we get
$\langle T_{\pm\pm}^{\rm bdy} \rangle_{\rm BD} =- (\kappa/ \pi) t_\pm (\sigma^\pm)= \kappa H^2/(4\pi)$
for $t_\pm (\sigma^\pm) =-H^2/4$.
In Eq.~(14), ignoring the bulk stress tensors, we obtain the Gibbons-Hawking temperature at the origin as $T_{\rm GH}=H/(2\pi)$, which is
the same as the result from the detector method where the observer is sitting at the origin \cite{Gibbons:1977mu,Spradlin:2001pw}.
Note that the boundary stress tensor is nothing but the vacuum expectation value of the normal ordered stress tensor,
{\it i.e.},
$\langle T_{\pm\pm}^{\rm bdy} \rangle =\langle :T_{\pm\pm}: \rangle$,
which is not covariant
\footnote[1]{Under the spacetime transformations of $x^\pm \to y^\pm = y^\pm (x^\pm)$,
the normal ordered stress tensor
 $\langle:T_{\pm\pm}:\rangle$ is anomalously transformed as
$\langle : T_{\pm\pm} (y^\pm) : \rangle \!=\! \left(\frac{\partial x^\pm}{\partial y^\pm}\right)^2 \langle : T_{\pm\pm} (x^\pm) : \rangle -\frac \kappa 2 \{x^\pm ,y^\pm\}$,
where the Schwarzian derivative is $\{x^\pm ,y^\pm\}=\frac{d^3 x^\pm}{d y^{\pm 3}} \big/  \frac{dx^\pm}{dy^\pm} - \frac 3 2 \left(\frac{d^2 x^\pm}{d y^{\pm 2}} \big/  \frac{dx^\pm}{dy^\pm}\right)^2$.
It breaks the general covariance.}
since the normal ordering responsible for a selection of modes is certainly associated with
a specific coordinate system \cite{doi:10.1142/p378}.
In addition, the boundary stress tensor cannot be  de Sitter invariant.
On the other hand, the stress tensor \eqref{eq:stress tensor pp mm} is true tensor
while the boundary stress tensor is anomalous under coordinate transformation.
The stress tensor \eqref{eq:stress tensor pp mm} and \eqref{eq:stress tensor pm} are compactly written in the de Sitter invariant form of $ \langle T_{\mu\nu} \rangle_{\rm BD} = \kappa H^2/(2 \pi) g_{\mu\nu}$
in the Bunch-Davies state,
which results in $\langle T_{\pm\pm} \rangle_{\rm BD}=0$ in the conformal gauge \eqref{eq:metric_lightcone}.
Using Eq.~\eqref{newcom},
we obtain the vanishing proper temperature.

\section{Conclusion and Discussion}
\label{sec:conclusion}
In conclusion, the two kinds of the proper temperatures of de Sitter space on the background of two-dimensional de Sitter space
warrant particular attention: one is for the de Sitter invariant Bunch-Davies
vacuum and the other is for the de Sitter non-invariant state.
The proper temperature in the former vacuum globally vanishes whereas the proper temperature in the latter state
becomes divergent at the cosmological horizon.
The above argument based on the background of de Sitter space was reexamined by
employing the two-dimensional dilaton gravity called the semiclassical Jackiw-Teitelboim gravity.
Interestingly, the dilaton parameter $\alpha$ can be related to the vacuum condition $t_\pm(\sigma)$,
so the dilaton configuration determines the state of the stress tensor.
For $\alpha=1$, the vacuum state turns out to be the de Sitter invariant Bunch-Davies vacuum: the proper temperature vanishes and
the dilaton field is finite. However, in a certain quantum state of $\alpha=2$,
the proper temperature is the observer-dependent and
a naked dilaton singularity occurs.
Consequently, the Bunch-Davies vacuum turns out to be the unique de Sitter invariant vacuum of
two-dimensional de Sitter space in equilibrium state and the proper temperature vanishes everywhere.

In fact, the Unruh's detector method
can be applied to de Sitter space \cite{Spradlin:2001pw}.
The thermal property can be deduced from the periodicity of thermal Green function for conformally invariant
scalar field propagating on de Sitter space.
In the original work by Unruh \cite{Unruh:1976db}, the observer is in flat space and he or she feels emitted thermal radiation
from the accelerated detector absorbing some energy from external accelerator.
There is a nice correspondence between Rindler space and de Sitter space.
A common feature to a uniformly accelerated observer in Minkowski space and
an observer at a fixed distance from the cosmological horizon
is that both observers have horizons which prevent them from seeing the whole of the
spacetime. Thus, one can expect a similar thermal interpretation in de Sitter space.
However, as compered to the Gibbons-Hawking temperature resorting to the detector method where the observer is sitting
at the origin in static coordinate system,
our investigation for the temperature
rests upon the Tolman's formulation relying on the covariant stress tensor approach.
Consequently, the proper temperature called the Tolman temperature
was read off from coordinate invariant quantities of the proper energy density and the trace of
the stress tensor.

The final comment is in order.
In formulating the proper temperature in thermal equilibrium \cite{Tolman:1930zza,Tolman:1930ona},
Tolman assumed that the stress tensor should be traceless; in other words, the equation of state should be satisfied.
The first law of thermodynamics consists of the proper energy density and the proper pressure,
in which one of them should be eliminated by using the equation
of state; however, in the presence of trace anomaly, eliminating one quantity leaves the trace term in the first law of thermodynamics.
This is the essential reason why the Stefan-Boltzmann law should be modified
such as Eq.~\eqref{eq:eff sb law} when the trace of the stress tensor does not
vanish \cite{Gim:2015era}. Quantum-mechanically radiating systems are commonly associated with
trace anomalies, so it would be interesting to study what happens in other gravitational systems
related to de Sitter space.

%\section{conclusion}
\label{sec:conclusion}

%Birrell-Davies \cite{birrell1984quantum}\\
%\textcolor{blue}{
%De Sitter space is the curved spacetime which has been most studied by quantum field theorist.
%The reason for this special attention stems from the fact that de Sitter space is the unique maximally
%symmetric curved spacetime.
%It enjoys the same degree of symmetry as Minkowski space (ten Killing vectors), which greatly
%facilitates technical computations as far as quantum field theory is concerned.
%Even so, the presence of curvature, and non-trivial global properties, introduce
%new aspects to the quantization of fields in de Sitter space.}

\acknowledgments
This work was supported by the National Research Foundation of Korea(NRF) grant funded by the Korea government(MSIT) (No. NRF-2022R1A2C1002894).
WK was partially supported by Basic Science Research Program through the National Research Foundation of Korea(NRF) funded by the Ministry of Education through the Center for Quantum Spacetime (CQUeST) of Sogang University (NRF-2020R1A6A1A03047877).
HE was partially supported by Basic Science Research Program (NRF-2022R1I1A1A01068833).\\

\textbf{Data Availability}\\
This manuscript has no associated data.

%%%%%%%%%%%%%%%%%%%%%%%%%%%%%%%%%%%%%%%%%%%%%%%%
%%%%%%%%%%%%%%%             References         %%%%%%%%%%%%%%%%

%\bibliographystyle{mybib}
%\bibliographystyle{apsrev4-1} % PRD
\bibliographystyle{JHEP}       %% JHEP.bst

\bibliography{references}

\end{document}